\documentstyle[12pt,aasms4]{article}
\begin{document}

\begin{center}
\title{Why Are The Secondary Stars in Polars So Normal?}

\author{Thomas E. Harrison}

\affil{Department of Astronomy New Mexico State University, Box 30001, MSC 4500, Las Cruces, NM 88003-8001}

\authoremail{ tharriso@nmsu.edu}

\author{Steve B. Howell}

\affil{WIYN Observatory and National Optical Astronomy Observatories, 950 North Cherry Avenue, Tucson, AZ 85726}

\authoremail{howell@noao.edu}
 
\author{Paula Szkody}

\affil{Department of Astronomy, University of Washington, Box 351580, Seattle, WA 98195}

and

\author{France A. Cordova}

\affil{Institute of Geophysics and Planetary Physics, Department of Physics,
University of California, Riverside, CA 92521}

\end{center}

\noindent
{\it Key words:} infrared: stars --- cataclysmic variables

\abstract{ We have used NIRSPEC on Keck II to obtain $K$-band spectroscopy of
several magnetic cataclysmic variables. These data reveal that the 
secondary stars in these binary systems have spectra that are consistent
with normal, late-type dwarfs in both their atomic and molecular line 
strengths, as well as in the slopes of their continuua. This result is in stark contrast to the
infrared spectra of their non-magnetic
cousins, nearly all of which show peculiar abundances, especially of CNO 
species and their isotopes. It appears that the evolutionary path taken by the
secondary stars in magnetic systems differs from that for the
non-magnetic systems. We discuss the implications of this result.}

\section{Introduction}

Cataclysmic variables (CVs) are short-period binary systems consisting of a 
white dwarf primary that is accreting material via Roche-lobe overflow from 
a low mass, late-type secondary star. CVs can be divided into two main classes:
magnetic and non-magnetic. In magnetic systems, the primary white dwarf
has a strong magnetic field (0.5 MG $\leq$ B $\leq$ 240 MG). In ``Polars'',
the magnetic field of the white dwarf is strong enough to capture the 
accretion stream close to the secondary star. In ``Intermediate Polars'' 
(IPs), the magnetic field of the white dwarf is believed to be weaker 
(B $\leq$ 8 MG), and is insufficient to completely
prevent the formation of an accretion disk. IPs therefore exhibit behavior
that can be unique, or common to both magnetic and non-magnetic classes.

The commonly proposed evolutionary history for cataclysmic variables has
been assumed to be similar for both the magnetic and non-magnetic systems
(King et al. 1994, Kolb 1995),
and has three main phases. First, the orbital separation of the wide binary of 
the pre-CV is rapidly shrunk in a common envelope phase where the
secondary star orbits inside the red giant photosphere of the white dwarf
progenitor. The second phase is a very long epoch where gravitational 
radiation, or a magnetically constrained wind from the secondary star extracts 
angular momentum from the binary (magnetic braking), resulting in the
eventual contact of the photosphere of the secondary star with its Roche lobe.
The final phase begins once the secondary star contacts its Roche lobe, mass 
transfer to the white dwarf is initiated, and all of the phenomena associated 
with CVs is observed. 

Based on the current paradigm (see Howell, Nelson, \& Rappaport 2001, and 
references therein) it can be argued that the majority of secondary stars in CVs
should show little signs of evolution at the time of contact---the duration of 
the CV formation process is a small fraction of a low mass star's main 
sequence lifetime. In addition, after contact, the secondary star begins to 
{\it lose mass},
extending its lifetime, and preventing it from ever forming a He-burning
core (Howell 2001).  Recent results, however,
challenge this idea. Infrared spectroscopy of two dozen CVs by Harrison
et al. (2004, 2005), and UV spectroscopy (G$\ddot{a}$nsicke et al. 2003,
and references therein) find evidence for peculiar abundance ratios in the 
secondary stars of non-magnetic CVs. The
dominant anomalies are deficits of carbon and enhancements of nitrogen. In 
addition, several CVs show evidence for enhanced levels of $^{\rm 13}$CO
in their $K$-band spectra. Taken
together, these results suggest that CNO-processed material has found its
way into the photospheres of CV secondary stars. Either the accretion of
material during the common envelope phase and/or from classical novae 
eruptions is much more efficient than expected, or the secondary stars in 
non-magnetic CV systems started out with much larger masses (to allow CNO 
burning) than current population synthesis theories predict. If magnetic 
systems followed the same evolutionary path prior
to the contact phase, then they too should have secondary stars with similar 
abundance anomalies.
In the following we present new infrared $K$-band spectroscopy for five 
polars: VV Pup, ST LMi, AR UMa, MR Ser and SDSS J1553+5516.  

\section{Observations}

Infrared spectroscopy for the program objects was obtained using 
NIRSPEC\footnote{For more on NIRSPEC go to http://www2.keck.hawaii.edu/inst/nirspec/nirspec.html} on Keck II in photometric conditions on 17 February 2005. A 
journal of our observations is presented in Table 1. We used NIRSPEC in 
low resolution mode with a 0.38" slit. The grating tilt was set so as to
cover the wavelength region 2.04 $\mu$m $\leq$ $\lambda$ $\leq$ 2.46 $\mu$m,
with a dispersion of 4.27 \AA ~pixel$^{\rm -1}$. We employed the two-nod
script, and we used four minute exposure times for all of the program CVs. To
correct for telluric absorption, we observed bright A0V stars located
close to the program objects so as to minimize their relative differences in 
airmass. These data were reduced using the
IDL routine ``$redspec$'', specially developed for NIRSPEC\footnote{Details
about the~$redspec$ package can be found here: http://www2.keck.hawaii.edu/inst/nirspec/redspec/index.html}.
In the $K$-band, the spectra of A0V stars are nearly featureless, except
for the prominent H I Brackett $\gamma$ feature at 2.16 $\mu$m. The $redspec$ 
package does not attempt to correct for this feature, but can interpolate
across such lines to reduce their impact upon division
into the program star spectrum. Note that there is a weak telluric feature
located very close to the Brackett $\gamma$ line, and thus the H I line 
profiles in spectra produced by the division of a ``patched'' A-star
spectrum are slightly compromised. 

In Fig. 1 we present the final, medianed spectra of VV Pup, ST LMi, AR UMa, 
and SDSS J1553+5516. The NIRSPEC spectrum of MR Ser is shown in Fig. 2, where
it is compared to a 2003 May spectrum obtained using SPEX on the IRTF. As seen 
in Table 1, the time spent on
ST LMi and AR UMa was relatively short, $\leq$ 10\% of an orbital
period, and thus smearing of the secondary star features due to orbital 
motion is not significant. This is not the case for VV Pup, SDSS J1553+5516, and 
MR Ser. For MR Ser and VV Pup we have used published ephemerides by Schwope 
et al. (1993) and Walker (1965) to correct for the radial velocity 
motion of their secondary stars. A radial velocity analysis of the VV Pup
data set presented here (Howell et al. 2005, in prep.), confirms the phasing
from Walker (1965). No such ephermerides exist for SDSS J1553+5516,
and thus the orbital motion of its secondary star cannot be accounted for. 
Fortunately, its spectrum does not show signs of significant smearing. 
The spectra in Fig. 1 have been smoothed to a resolution of 5.1 \AA 
~pixel$^{\rm -1}$ to allow us to directly compare them to spectra of late-type 
dwarfs obtained using SPEX on the IRTF (from Harrison et al.  2005). It is
important to note that the $redspec$  package flux calibrates the spectra
using a blackbody spectrum while the SPEXTOOL package (Vacca et al. 2003) used 
to produce the late-type dwarf spectra from SPEX data employs a model 
A-star atmosphere. Any subtle differences in the slopes of the continuua
between the program objects presented here, and the late-type templates
observed with SPEX, could in fact be due to the slightly different flux 
calibration procedures used by the two packages, and not due to any 
intrinsic differences.

\section{Results}

Close examination of the spectra shown in Fig. 1 reveal no significant
abundance anomalies (see Harrison et al. 2005 for identifications of the
stronger spectral features found in late-type dwarfs). The atomic absorption
lines and CO absorption features appear 
to have strengths that are consistent with those expected for late type 
dwarfs. We have derived spectral types for each of the secondary stars and 
list them in Table 1. The spectral features in VV Pup are somewhat broader
and/or weaker than in the other program
objects, suggesting a large rotational velocity, or improper phasing. VV Pup
was also the faintest of the five polars, and thus its spectrum is somewhat
noisier.  Even with this, it is clear that the Na I
doublet (near 2.20 $\mu$m) in VV Pup has a strength relative to the CO features
(at 2.294 $\mu$m, and redward) that is consistent with it being a normal,
very late M-type dwarf.
As demonstrated by the weakness of their emission lines, all of
the polars (except MR Ser) were in low states, a time when the accretion 
rate from the secondary star drops to a very low value, and the activity
level of the system declines. As seen in Fig. 2, the emission lines from He I 
and H I in MR Ser were 
prominent, suggesting it was in an active state, though no visual estimates
of the system brightness exist to confirm this conclusion. Due to this activity
level, the spectrum of the secondary star in MR Ser is considerably diluted
compared to that observed during a lower level of activity.  
During the low state, analysis of the IR spectrum of MR Ser obtained with 
SPEX on the IRTF lead to a classification of 
M8V. Thus, the
secondary stars of these five polars, plus that of the proto-type system, AM 
Her (Harrison et al.  2005), appear to be relatively normal late-type dwarfs.

That the secondary stars of polars show no evidence for peculiar abundance
patterns is in stark contrast to our results for non-magnetic systems 
(Harrison et al. 2004, 2005). In those efforts, only a single system
(IP Peg) out of twenty CVs had a secondary star with normal CO absorption 
features. {\it This suggests that the evolutionary histories of magnetic and 
non-magnetic systems are different. }

Since all known CV secondary stars have masses 
below 1.3 M$_{\sun}$, generating peculiar abundances via the CNO cycle in CV
secondary stars cannot occur after contact (Howell 2001). Thus, any observed 
abundance anomalies are either 1) the result of normal stellar evolution 
during the pre-CV phase, 2) are due to material of peculiar composition being 
accreted during the common envelope phase, or 3) comes from the accretion of 
classical novae ejecta once the mass transfer phase ensues (see Marks \& Sarna, 
1998). Both magnetic (e.g., V1500 Cyg, Stockman et al. 1988; CP Pup, Diaz \& 
Steiner 1991; V2214 Oph, Baptista et al.  1993; V2487 Oph, Hernanz \& Gloria 
2002;  BT Mon, White et al. 1996; GQ Mus, Diaz \& Steiner 1994) and 
non-magnetic systems have been observed as classical novae. As shown in
calculations by Livio, Shankar, \& Truran (1988), the field strengths found
in polars (and expected in IPs) are insufficient to inhibit, or dramatically
alter, classical novae outbursts. Thus, the accretion of novae ejecta as an 
explanation for unusual 
abundance patterns in CV secondary stars appears to be eliminated. It
is critical, however, to show that one or more of the classical novae listed
above are true polars, and are not simply IPs like the proto-type magnetic 
classical nova DQ Her. 

Kolb (1995), King et al. (1994), and Leibert et al. (2005) all conclude that 
the presence of a 
highly magnetic ``white dwarf'' primary cannot strongly affect the common 
envelope phase of evolution unless the secondary star orbits close to the 
magnetospheric radius of the magnetic core. If true, we are left to conclude 
(under the standard CV evolutionary paradigm) that the abundance anomalies we 
detect in non-magnetic systems must be due to evolutionary effects in the 
secondary stars themselves. This implies that the secondary stars of 
non-magnetic systems must start life with a mass sufficient to initiate the 
CNO cycle (M $\geq$ 1.3 M$_{\sun}$) so as to enrich some layers within their 
atmospheres. The current, low mass secondary star is then the stripped remains 
of this more massive object.

If peculiar CNO abundances in non-magnetic CVs result from
normal stellar evolution, then the secondary stars in polars must have 
entered and exited the pre-contact phases with masses similar to what 
is currently observed.  Given the observed preference of polars to have
shorter periods than non-magnetic systems (see King et al. 1994, and 
references therein), it is relevant to ask whether all CVs below the famous
period gap are simply the product of pre-CV binaries with initially low mass 
secondary stars. The UV study by G$\ddot{a}$nsicke et al. (2003) shows that at
least one, normal, non-magnetic system below the priod gap, BZ UMa, shows a 
large N V/C IV emission line ratio. It is vital to attempt to obtain data
on more CV systems below the period gap to investigate whether this is true
for all short period non-magnetic systems.

A recent FUV survey of eleven polars by Araujo-Betancor et al. (2005) 
finds normal N~V/C~IV emission line ratios for several systems, 
indicating that the material being transfered to the white dwarf in those 
polars is not of unusual composition, adding further weight to our results.
We note, however, that the {\it asynchronous}, long period polar, BY Cam, has 
extreme N V/C IV line ratios (Mouchet et al. 2003), providing at least one, 
albeit peculiar, counter-example. The strong correspondence between the 
N V/C IV emission line ratios in the UV, and the strength of the secondary 
star's CO features in the IR, is remarkable, and it
appears that one can be used as a proxy for the other. 

We conclude that the evolutionary history of most polars appears to differ from 
that of the majority of non-magnetic CVs. While the exact origin for this 
difference could
be the initial masses of their secondary stars, it is hard to understand
why low mass binaries preferentially go on to produce magnetic white dwarf
primaries. A related conundrum is the apparent lack of ``pre-polars''
(Schmidt et al. 2005, Wellhouse et al. 2005, and references therein), binary 
systems containing magnetic white dwarf primaries and main sequence secondary 
stars that are not in contact with their Roche lobes. Does the magnetism
of the white dwarf primary accelerate the pre-contact evolution to the
point that nearly all polars are born in their currently observed
states? Schmidt et al. suggest 
that the small family of ``Low Accretion Rate polars'' (LARPs, Schwope et al. 
2002), including SDSS J1553+5516, are in fact the
pre-polars. However, EF Eri is currently behaving in a fashion that is
{\it similar} to this group of LARPs, a condition attained only seven years 
after an observed high state (Harrison et al. 2004). 
Given the ability of EF Eri to get stuck in such a prolonged low state, it is 
probably too early to conclude that the recently discovered LARPs never have 
high states.

Our results for magnetic systems has shed new light on non-magnetic
systems: It now seems that the most likely path for the extreme levels of 
carbon depletion found in non-magnetic systems is the pre-contact evolution of 
the secondary star. The secondary stars in most non-magnetic CVs 
must have started out life with much higher masses than is observed now, in
direct conflict with population synthesis theories. 

Given this scenario, it is interesting to
postulate whether Algols might be the progenitors of CVs. Algols are plentiful,
both components in most systems appear to have had initial masses high enough 
to ignite the CNO cycle during their lifetimes, and many Algols have orbital 
periods ($\sim$ 1 day) that fall in the range to make suitable pre-CV 
candidates. In addition, the stellar components in numerous Algol systems show 
peculiar abundances of CNO elements, including enhanced levels of nitrogen, and
deficits of carbon (Cugier \& Hardorp 1988, Cugier 1989, Parthasarathy et al. 
1983, Tomkin \& Lambert 1989). Two particularly relevant Algols, that already
have total system masses similar to those of the longest period CVs, are TT Hya
and S Cnc. Both TT Hya and S Cnc have B9.5V primaries, and cool late-type 
giant/subgiant secondary stars with exceptionally small masses of 0.4 and 
0.2 M$_{\sun}$, respectively (Olson \& Etzel 1993; Etzel 1988). The 
existence of a ``flip-flop'', where 
the primary and secondary star switch roles, was long ago (Paczynski 1971) 
proposed to explain the presence of low mass, evolved secondary
stars in Algols with high mass primaries. It might be time to explore the 
possibility that a subset of Algols are the progenitors of non-magnetic CVs,
where the peculiar secondary stars in both are the remnant ``cores'' of the
same, once more massive stars.

~
\noindent
Acknowledgments: Data presented herein were obtained at the W.M. Keck 
Observatory, which is operated as a scientific partnership among the 
California Institute of Technology, the University of
California and the National Aeronautics and Space Administration. The 
Observatory was made
possible by the generous financial support of the W.M. Keck Foundation.
The authors wish to recognize and acknowledge the very significant cultural 
role and
reverence that the summit of Mauna Kea has always had within the indigenous 
Hawaiian community.  We are most fortunate to have the opportunity to conduct 
observations from this mountain.

\begin{flushleft}
{\bf References}

Araujo-Betancor, S., G$\ddot{a}$nsicke, B. T., Lonk, K. S., Beuermann, K., de Martino, D., Sion, E. M., \& Szkody, P. 2005, ApJ, 622, 589

Baptista, R., Jablonski, F. J., Cieslinski, D., \& Steiner, J. E., 1993, ApJ, 406, L67

Cugier, H. 1989, A\&A, 214, 168

Cugier, H., \& Hardorp, J. 1988, A\&A, 202, 101

Diaz, M. P., \& Steiner, J. E. 1991, PASP, 103, 964

Diaz, M. P., \& Steiner, J. E. 1994, AJ, 425, 252

Etzel, P. B. 1988, AJ, 95, 1204

G$\ddot{a}$nsicke, B. T., Szkody, P., de Martino, D., Beuermann, K., Long, K. S.,
 Sion, E. M., Knigge, C., Marsh, T., Hubeny, I. 2003, ApJ, 594, 443

Harrison, T. E., Osborne, H. L., \& Howell, S. B. 2005, AJ, 129, 2400

Harrison, T. E., Osborne, H. L., \& Howell, S. B. 2004, AJ, 127, 3943

Hernanz, M., \& Sala, G. 2002, Science, 298, 393

Howell, S. B., Harrison, T. E., Szkody, P. 2004, ApJ, 602, L49

Howell, S. B., Nelson, L. A., Rappaport, S. 2001, ApJ, 550, 897

Howell, S. B. 2001, PASJ, 53, 675

King, A. R., Kolb, U., de Kool, M., \& Ritter, H. 1994, MNRAS, 269, 907

Kolb, U. 1995, in ``Cape Workshop on Magnetic Cataclysmic Variables'', ASP
Conf. Series, Vol 85, ed. D. A. H. Buckley and B. Warner (ASP: San Francisco),
p 440

Leibert, J., et al. 2005, AJ, 129, 2376

Livio, M., Shankar, A., \& Truran, J. W. 1988, ApJ, 330, 264

Marks, P. B., \& Sarna, M. J. 1998, MNRAS, 301, 699

Mouchet, M., et al. 2003, A\&A, 401, 1071

Olson, E. C., \& Etzel, P. B. 1993, AJ, 106, 1162

Paczynski, B. 1971, Ann. Rev. Astron. Astrophys. 9, 183

Parthasarathy, M., Lambert, D. L., \& Tomkin, J. 1983, MNRAS, 203, 1063

Schmidt, G. D., et al. 2005, ApJ in press, astro-ph/0505385

Schwope, A. D., Beuermann, K., Jordan, S., \& Thomas, H. -C. 1993, MNRAS 278,
487

Schwope, A. D., Brunner, H., Hambaryan, V., Schwarz, R., Staude, A., Szokoly,
G., \& Hagen, H.-J. 2002, in ``The Physics of Catclysmic Variables and 
Related Objects'', ASP Conf. Ser. 261, ed. G$\ddot{a}$nsicke, B. T.,
Beuermann, K., \& Reinsch, K. (San Francisco: ASP), 102

Stockman, H. S., Schmidt, G. D., \& Lamb, D. Q. 1988, ApJ, 332, 282

Tompkin, J., \& Lambert, D. L., 1989, MNRAS, 241, 777

Vacca, W. D., Cushing, M. C., \& Rayner, J. T. 2003, PASP, 115, 389

Walker, M. F. 1965, Mitt. Sternwarte Budapest 57, 1

Wellhouse, J. W., Hoard, D. W., Howell, S. B., Wachter, S., \& Esin, A. A. 
2005, PASP, in press.

White, J. C., Schlegel, E. M., \& Honeycutt, R. K. 1996, ApJ, 456, 777
\end{flushleft}

\begin{deluxetable}{cccccc}
\tablecaption{Observation Journal}
\tablehead{\colhead{Object} &\colhead{P$_{\rm orb}$ (hr)} & \colhead{Number of Exps.}
&\colhead{Start (UT)}  &  \colhead{Stop (UT)} &\colhead{Spectral Type}
}
\startdata
VV Pup  & 1.674 & 22 &  08:23 &  10:04 & M7\\
AR UMa  & 1.932 &  6 &  10:23 &  10:50 & M5.5\\
ST LMi  & 1.898 &  6 &  10:56 &  11:23 & M6\\
SDSS J1553& 4.39  & 12 &  13:27 &  14:24 & M4.5\\
MR Ser  & 1.891 & 12 &  15:06 &  16:00 & M8\\
\enddata 
\end{deluxetable}

\begin{figure}
\plotone{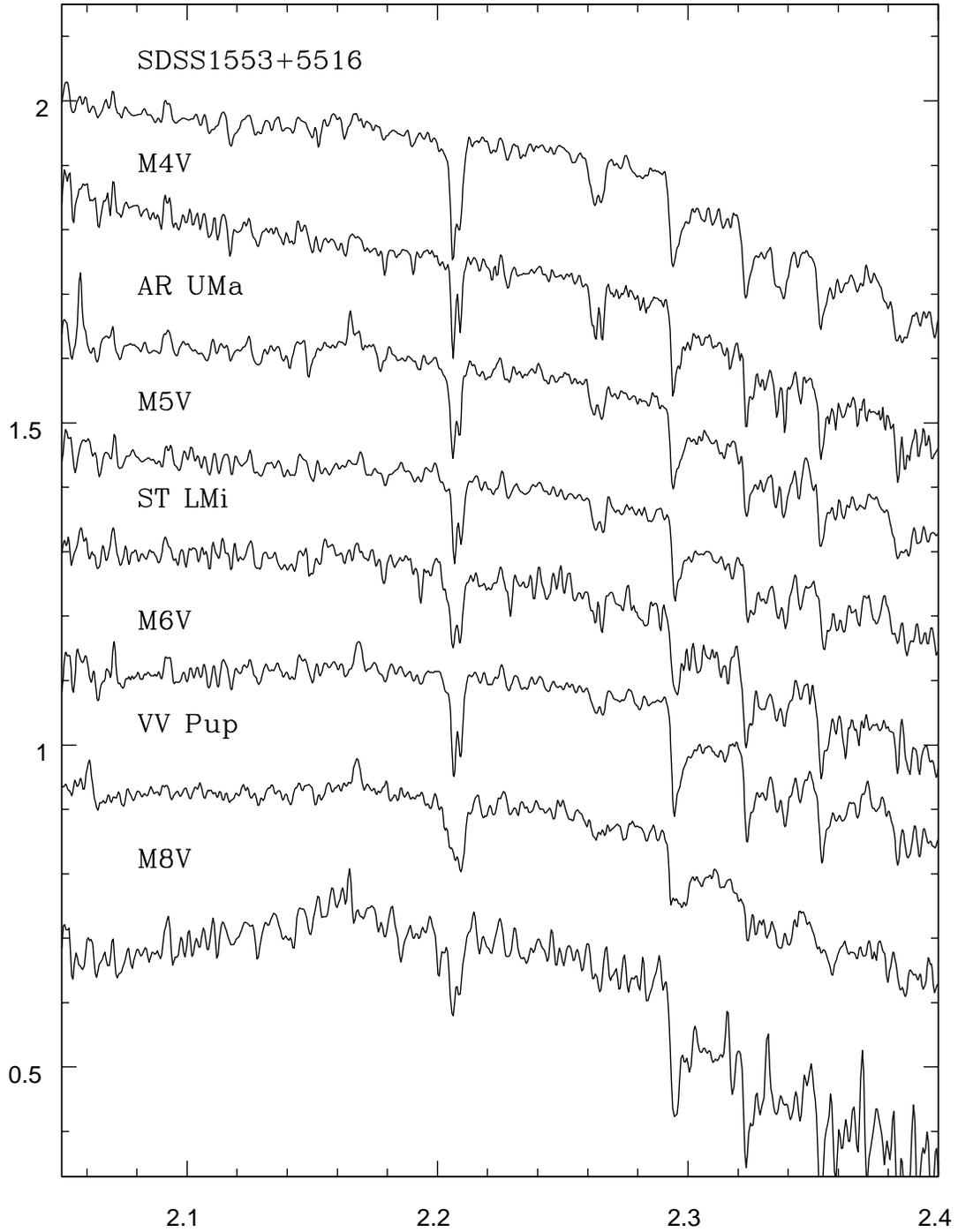}
\caption{The program object spectra compared to the spectra of some late-type
dwarfs. To provide a slightly more realistic match to the program object
spectra, the data for the M5V, M6V, and M8V have been rotationally broadened to
150 km s$^{\rm -1}$, while the M4V spectrum has been broadened to 110 km s$^{\rm -1}$.}
\end{figure}

\begin{figure}
\plotone{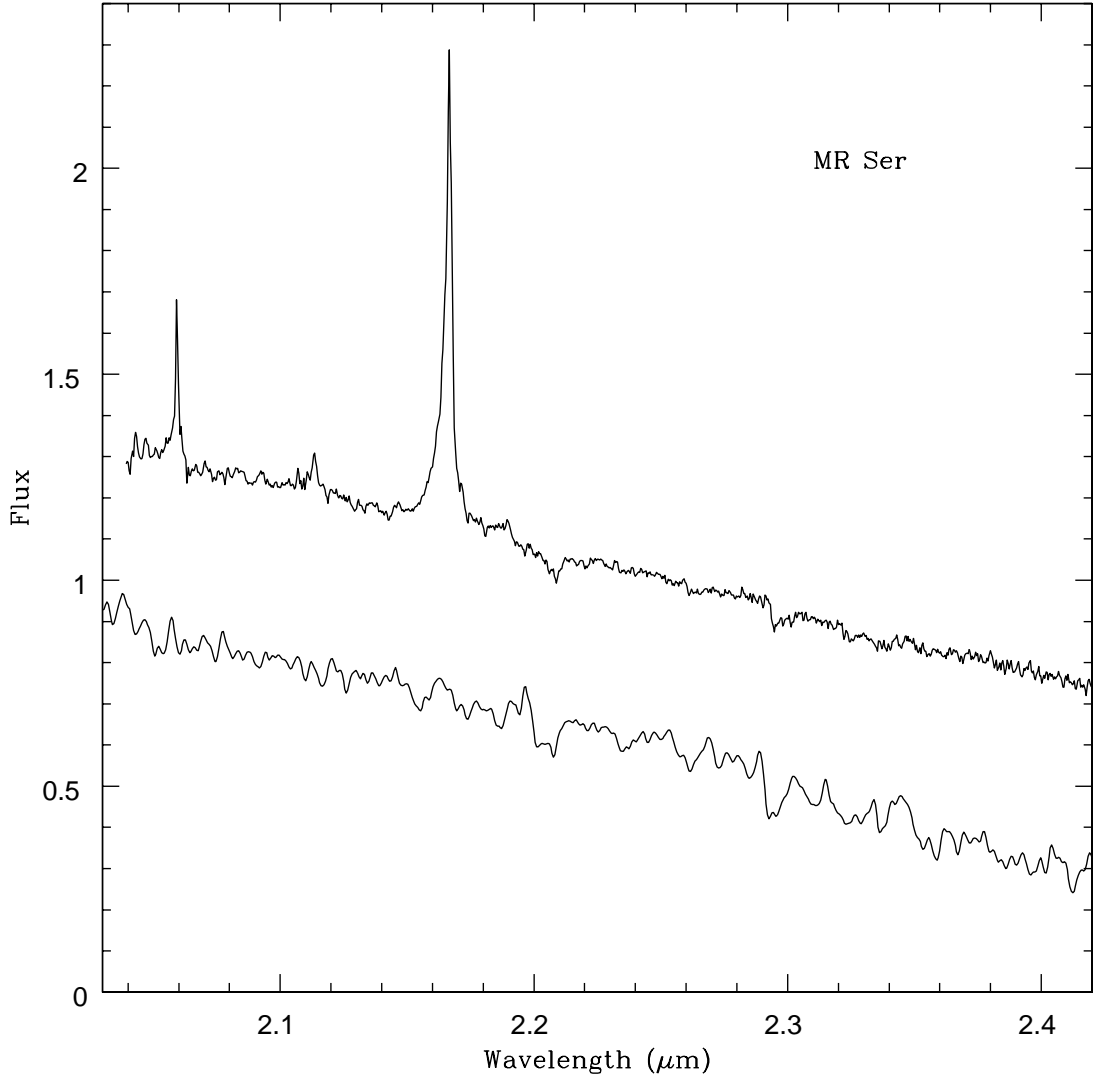}
\caption{The NIRSPEC data for MR Ser (top), compared with the IRTF + SPEX
spectrum obtained in 2003 May. Emission lines from He I at 2.06 and 2.11 $\mu$m,
and H I Br$\gamma$ at 2.16 $\mu$m are present in the NIRSPEC data, but
not in the SPEX data. Clearly, MR Ser was in a ``high state'' in 2005 February.
The NIRSPEC data have been Gaussian smoothed to a resolution of 5.1 \AA ~pixel$^{\rm -1}$, 
while
the SPEX data have been smoothed to a resolution of 40  \AA ~pixel$^{\rm -1}$.}
\end{figure}

\end{document}